\documentclass[sigconf]{acmart}

\AtBeginDocument{%
  }

\copyrightyear{2023}
\acmYear{2023}
\setcopyright{rightsretained}
\acmConference[CHI EA '23]{Extended Abstracts of the 2023 CHI Conference on Human Factors in Computing Systems}{April 23--28, 2023}{Hamburg, Germany}
\acmBooktitle{Extended Abstracts of the 2023 CHI Conference on Human Factors in Computing Systems (CHI EA '23), April 23--28, 2023, Hamburg, Germany}\acmDOI{10.1145/3544549.3585657}
\acmISBN{978-1-4503-9422-2/23/04}

\begin{document}

%%
%% The "title" command has an optional parameter,
%% allowing the author to define a "short title" to be used in page headers.
%\title{I Am Not Worried (Yet): Creative Professionals' Opinions and Expectations about Generative AI}
\title{Designing Participatory AI: Creative Professionals' Worries and Expectations about Generative AI}

%%
%% The "author" command and its associated commands are used to define
%% the authors and their affiliations.
%% Of note is the shared affiliation of the first two authors, and the
%% "authornote" and "authornotemark" commands
%% used to denote shared contribution to the research.

\author{Nanna Inie}
\email{nans@itu.dk}
\affiliation{%
  \institution{Center for Computing Education Research (CCER), IT University of Copenhagen}
  \streetaddress{Rued Langgaards Vej 7}
  \city{Copenhagen}
  \country{Denmark}
  \postcode{2300}
}

\author{Jeanette Falk}
\email{jeanette.falk-olesen@plus.ac.at}
\affiliation{%
  \institution{Dep. of Artificial Intelligence and Human Interfaces, University of Salzburg}
  \streetaddress{Jakob-Haringer-Straße 8}
  \city{Salzburg}
  \country{Austria}
  \postcode{5020}
}

\author{Steven Tanimoto}
\email{tanimoto@uw.edu}
\affiliation{%
  \institution{Paul G. Allen School of Computer Science \& Engineering, University of Washington}
  \streetaddress{1410 NE Campus Parkway}
  \city{Seattle}
  \country{USA}
}

%%
%% By default, the full list of authors will be used in the page
%% headers. Often, this list is too long, and will overlap
%% other information printed in the page headers. This command allows
%% the author to define a more concise list
%% of authors' names for this purpose.
%\renewcommand{\shortauthors}{Anonymous et al.}

%%
%% The abstract is a short summary of the work to be presented in the
%% article.

\begin{abstract}
Generative AI, i.e., the group of technologies that  automatically generate visual or written content based on text prompts, has undergone a leap in complexity and become widely available within just a few years. Such technologies potentially introduce a massive disruption to creative fields. This paper presents the results of a qualitative survey ($N$ = 23) investigating how creative professionals think about generative AI. The results show that the advancement of these AI models prompts important reflections on what defines creativity and how creatives imagine using AI to support their workflows.
Based on these reflections, we discuss how we might design \textit{participatory AI} in the domain of creative expertise with the goal of empowering creative professionals in their present and future coexistence with AI.
\end{abstract}

%%
%% The code below is generated by the tool at http://dl.acm.org/ccs.cfm.
%% Please copy and paste the code instead of the example below.
%%
\begin{CCSXML}
<ccs2012>
   <concept>
       <concept_id>10010147.10010178.10010216</concept_id>
       <concept_desc>Computing methodologies~Philosophical/theoretical foundations of artificial intelligence</concept_desc>
       <concept_significance>300</concept_significance>
       </concept>
   <concept>
       <concept_id>10003120.10003121.10011748</concept_id>
       <concept_desc>Human-centered computing~Empirical studies in HCI</concept_desc>
       <concept_significance>500</concept_significance>
       </concept>
   <concept>
       <concept_id>10003120.10003121.10003122</concept_id>
       <concept_desc>Human-centered computing~HCI design and evaluation methods</concept_desc>
       <concept_significance>300</concept_significance>
       </concept>
 </ccs2012>
\end{CCSXML}

\ccsdesc[300]{Computing methodologies~Philosophical/theoretical foundations of artificial intelligence}
\ccsdesc[500]{Human-centered computing~Empirical studies in HCI}
\ccsdesc[300]{Human-centered computing~HCI design and evaluation methods}

%%
%% Keywords. The author(s) should pick words that accurately describe
%% the work being presented. Separate the keywords with commas.
\keywords{participatory AI, participatory design, generative AI, creative professionals, creativity support}

%\received{20 February 2007}
%\received[revised]{12 March 2009}
%\received[accepted]{5 June 2009}

%%
%% This command processes the author and affiliation and title
%% information and builds the first part of the formatted document.
\settopmatter{printfolios=true}
\maketitle

\section{Introduction and background}
%Inkluder referencer fra R3

Recent developments in \textit{generative AI}, i.e., AI technologies that automatically generate visual or written content based on text prompts, have led to much speculation and concern  about what these developments may mean for different professions in the future, particularly for professionals where creativity accounts for a sizable part of their everyday work \cite{Press2023}. 
%\sandy{Any citations for preceding contention?}
Potential ``threats'' that generative AI models may pose for creative professionals (``creatives'') include the ability to automate generation of high(er) quality content (text, code, images, and video), increased content variety, and personalized content based on preferences of individual users and consumers. Some current discourse about generative AI models frame them as threatening the ownership and agency of creatives. See, e.g., \cite{cnnTheseArtists,waxyInvasiveDiffusion,Heikkila20222} for interviews with artists whose work was --- unbeknownst to them --- used to train AI models that generated images in the style of the artists' work. 
Rogers critically discusses this scenario in terms of `the attribution problem with generative AI' \cite{Rogers_2022_attribution}. 

%One response to this challenge is finding ways to safeguard creative work for being used as training material for generative AI models, ranging from embedding `watermarks' into their digital art (see e.g. \cite{zhang2018protecting}) to filing class-action lawsuits (see e.g. \cite{stablediffusionlitigationStableDiffusion}).
 
Conversely, other creatives express curiosity and excitement about the potential this technology may offer, e.g., \cite{creatoreconomyWillHelp}. 
% brief review of findings from user studies of recent AI-based creativity support tools, especially those that include some questions about the user’s attitudes toward human-AI collaboration via a particular tool or in a particular creative context. (R3)
Regardless of whether generative AI is seen as a blessing or curse, it is both timely and of research value to answer questions about how it and creatives can most fruitfully coexist.

One response to perceived threats posed by AI is the notion of \textit{participatory AI}, where the goal is to include `wider publics' in the development and deployment of AI systems \cite{birhane2022power}. 
Historically, the emergence of \textit{participatory design} (PD) in the 1970s was motivated by efforts to rebalance \textit{``power and agency in the professional realm''} \cite{bannon2018reimagining} in order to 
empower workers to \textit{``codetermine the development of the information system and of their workplace''} \cite{clement1993retrospective}.
%In one of the first reviews of participatory design (PD) projects, Clement and van den Besselaar position PD as historically concerned with \textit{``the empowerment of workers so that they can codetermine the development of information systems and their workplace''}. 
%Described as essential to the notion of `user participation' is \textit{``the right of people to have a direct influence on matters that concern them in their work''}\cite{clement1993retrospective}. 
In light of this historical backdrop, participatory AI is expected to empower those affected by the development of novel technologies by enforcing values of inclusion, plurality, collective safety, and ownership \cite{birhane2022power}.

\looseness-1 One step in the direction of participatory AI is to understand the needs of people and communities affected. %Include R3 references here?
Related research in this direction includes Singh et al., who explored which assumptions and expectations creative writers have for a supporting AI tool \cite{singh2022hide}; Guzdial et al., who explored designers' expectations for AI-driven game-level editors \cite{guzdial2019friend}; and Zhu et al., who argued for a better understanding of game designers' needs when co-creating with AI \cite{zhu2018explainable}.

Our paper contributes to this objective by surfacing and categorizing concerns and expectations that creatives of different types currently have about the effect of generative AI on their work. 
It represents the authors'
%\sandy{Re above, by "our" do you mean your own or the field's? Perhaps you should say "the"?}
first empirical research into the question of \textbf{How might we design and perform \textit{participatory AI?}} 
%In this instance, we examine generative AI used in creative fields. 
This question is particularly relevant to those who design and develop AIs for creatives and those who design and develop creativity support tools that use AI technology.

\looseness-1 Our findings, albeit preliminary, identify important topics that may inform participatory design of generative AI so creatives can \textit{``influence digital technologies that will change their work practices or everyday life''} \cite{bodker2022participatory}, the goal of participatory design in its essence. Our contributions include the following. (1) We introduce new conceptions about what constitutes creativity in relation to generative AI. (2) We categorize some reasons why creatives are and are \textit{not} concerned about novel generative AI. (3) We categorize reasons why some creatives are curious and excited about AI and how it might augment their creative processes. (4) We discuss possible foci for the design of participatory AI aimed at helping creative professionals \textit{Understand} AI, \textit{Cope} with AI, \textit{Adapt} to AI, and \textit{Exploit} AI.

%\begin{enumerate}
   %\item We introduce some new conceptions of what constitutes creativity in relation to generative AI.
   % \item We categorize some of the worries of creative professionals towards generative AI, as well as reasons for \textit{not} worrying about novel generative AIs.
    %\item We categorize reasons that creative professionals are excited about AI and how it may help their creative processes.
    %\item We discuss possible foci for the design of participatory AI aimed at helping creative professionals \textit{Understand} AI, \textit{Cope} with AI, \textit{Adapt} to AI, and \textit{Exploit} AI.
%\end{enumerate}

\vspace{-0.5em}
\section{Methods}
%Inkluder nogle af R1's forslag til at uddybe metodevalg
We conducted a qualitative survey with open-ended questions designed to encourage longer answers and reflection. The survey format let respondents participate asynchronously, while allowing us to discover themes and directions for further in-depth research.
%Some justification for survey design: F.ex:  why did the authors choose a qual survey instead of an in-depth interview,
The survey was circulated to the authors' networks of creative professionals as well as on social media. The call was posted as an open question of `Are you a creative professional/professional creative, and do you have opinions about generative AI that you would like to share with us?' The term `creative' was left to self-definition, and we asked the respondents to explain the role of creativity in their profession. We collected responses over a period of approximately two months in late 2022. We offered a draw of five \$25 gift cards to Amazon as symbolic compensation for participation. The study and survey were approved by the ethical committees of the authors' universities.

\vspace{-0.5em}
\subsection{Participants}

We received 23 responses to the survey from creatives residing in Denmark (10), Germany (4), the United Kingdom (4), USA (3),  Turkey (1), and Morocco (1). Respondents were between 21 and 55 years old, distributed as 21-25 (4), 26-30 (1), 31-35 (8), 36-40 (3), 41-45 (6), and 51-55 (1). 10 respondents identified as female, 12 as male, and 1 as non-binary. The respondents worked in a variety of fields, from computer science research to design of UX/UI and games to teaching. Most respondents came from software-oriented creative domains, and our findings should be read with this limitation in mind (see Section \ref{conclusion} for a discussion of this limitation).
%A comment that most came from a smaller number of creative fields (CS and games) --> Might entail a pro-AI perspective
We were more interested in people self-qualifying as a ``creative professional'' where creativity plays a significant role in their work, than we were in specific job titles. The responses, as well as a detailed overview of respondents, are presented in the supplementary material.

\vspace{-0.5em}
\subsection{Survey and analysis}
The survey consisted of both demographic questions and six questions related to our research interest (which we list below). We designed the questions to elicit respondents' general understanding of and attitudes towards AI and creativity. We sought to prompt a deeper level of reflection and tried to avoid overloading respondents with questions.

\begin{enumerate}
    \item In your own words, how would you define what AI (Artificial Intelligence) is?
    \item Do you believe computers can be creative? Why/why not?
    \item A standard definition of a creative idea is that it is: 1. original (new, either to the creator or to human history in general), 2. useful (in some context), and 3. surprising (it seems unlikely but possible). Given this definition, do you believe a computer/an AI algorithm can generate creative ideas? Why/why not?\footnote{This definition is a compilation of three-criterion definitions by, e.g., Boden \cite{boden2004creative} and Simonton \cite{simonton2012taking}.}
    \item Are you excited about AI contributing to creative work in your profession? Why/why not?
    \item Do you worry about AI replacing creative work in your profession? Why/why not?
    \item Which role do you think AI will play in your profession in the near and far future?
\end{enumerate}

\noindent 
Questions (4) and (5) were swapped for approximately half the respondents (in two different instances of the survey) to avoid priming respondents in any specific direction.

We performed a thematic analysis as described by Braun and Clarke \cite{braun2012thematic} on the responses. We tagged responses individually with different codes and then clustered them into sub-themes, which we highlight in bold throughout Section 3. 

%Or the authors did not consider such contextualization as an important factor in designing the survey. In this sense, the findings now read relatively more about the general understanding of generative AI, in other words, not distilling an in-depth understanding.

\vspace{-0.5em}
\section{Survey responses}
\vspace{-0.2em}
\subsection{How intelligent or creative is generative AI?}
In order to inform the design of participatory AI, it is relevant to understand how creatives currently conceive of AI and its limits. These factors can inform decisions about how to design participation processes to, for instance, include more or less information and discussion about the state of AI.
%should be more tightly related to RQ - how can we strengthen transition between analysis and discussion, to make the findings more actionable and accessible? R1

\subsubsection{What is AI?}

Answers to the question ``In your own words, how would you define AI?'' varied, especially on two scales: \textbf{technical depth} (from superficial to deep understanding) and \textbf{agency of AI} (from no agency to high agency). In terms of \textbf{technical depth}, some respondents, naturally, had a deeper understanding of AI algorithms than others, e.g., from \textit{``... digital solutions that are trained to be helpful in specific ways''} (P21) (technically superficial) to \textit{``A system capable of making dynamic choices based on input, dynamic as in non-binary evaluation of input referencing data model, a model which would ideally evolve through feedback of external verification of multiple processes'} (P2) (technically advanced).

We also saw interesting variation in the level of \textbf{agency} ascribed to the AI system, from no agency at all: \textit{``AI is a set of rules, defined by humans, which a computer can follow.''} (P3) to a high degree of agency: \textit{``it's a computer that over time improves itself in the tasks it has to solve by collecting information and inputs from humans''} (P7). These understandings may influence creatives' judgments of the degree to which AI can support them and contribute to/replace tasks in their creative processes.

 We tagged 7 responses as portraying a relatively deep technical understanding with no agency to the computer. Six responses were tagged with a more superficial technical understanding and no agency ascribed to the computer. 10 responses were tagged as superficial technical understanding with a high degree of computational agency, and no responses were tagged as deep technical understanding and high agency of the computer. An overview is shown in the supplementary material, Figure \ref{fig:whatisai}.
%\sandy{Comment on the preceding results and/or show them in a graph form (with 4 quadrants)?}

\subsubsection{New definitions of creativity} \label{creativitydefinitions}
The presence of generative AI encourages us to reevaluate and question our understanding of creativity and creative ideas. Most respondents who denied that AIs can be considered creative disputed the computer's capacity to generate \textit{original} output since it is trained only on already existing (human) input. However, one respondent wrote in answer to ``Do you believe computers can be creative?'': \textit{``I kind of resent it - but yeah. If creativity is defined as something useful and new, then yeah I think so. Even though AI’s [sic] rely on training data and existing man-made patterns (which some might use to criticize AI’s as being derivative or as simply reproducing what already exists) the process of combining stuff into a new “something” isn’t really THAT different from what humans do… it’s just bigger in scale and I guess you might argue that humans are also just “trained on” a bunch of data… we also carry around a repertoire of input we can draw on to come up with ideas [...] ideas are always rooted in some pre-existing thing(s)''} (P5). Another participant noted that \textit{``What is my brain if not a computer that takes in all this provided data and produces its own result from a mix of the inputs? If that result is `creative', then why is an AI not?''} (P20). 

This understanding is consistent with a traditional definition of creative ideas as being\textbf{ `novel', `useful', and `surprising'}, e.g., \cite{boden2004creative, simonton2012taking}. However, one respondent noted that \textit{``Computers aren't creative by themselves as they only follow the orders that someone gives them''} (P6). In P6's understanding, creativity entails \textbf{agency} or \textbf{initiative}, which is not historically a property of the three-criterion definition of creativity. 

\textbf{Intention} and \textbf{sentience} were described as criteria for creativity by some respondents: \textit{``There is not intention''} (P1), \textit{``Creativity stems from personal experiences/knowledge/emotions and the need to express/communicate/use this [...] Creativity lies not in the creation, but in why we create. Programs can emulate this, but without true sentience, it will always be [an] emulation''} (P10), and %\textit{``The result, though, can be seen as creative by humans - but the computer doesn't really know or care''} (P21). 
\textit{``The computer still isn't creative, it's still just doing what it's told [...] Maybe I think it needs feelings to be truly creative?''} (P23).

Other conditions for creativity were also evoked in the answers, such as \textbf{(self-)awareness}: \textit{``I think that true creativity requires a sense of self and self-awareness''} (P8). 
\textit{``They are not creative in themselves; they are producing content unaware of the value they just created''} (P21). Even \textbf{experiences} and \textbf{inspiration} were evoked: \textit{``Computers can solve problems and create art and everything, but it will all be logic and calculated and not because it got a sudden burst of inspiration or remembered something that happened in the second grade''} (P23).

\looseness-1 Even if we do not assume that these definitions should be unanimously integrated into a scholarly or theoretical definition of creativity, it is interesting that reflecting on creativity in relation to the role of generative AI raises different conceptions of what creativity entails. 

\vspace{-1em}

\subsection{I Am Not Worried (Yet)}
\vspace{-0.2em}
%\sandy{In the following, you note that three respondents answered yes but provide answers only for two; further, I'm not sure what the second ("uncanny") really means.}
Only three of our 23 respondents unambiguously answered yes to being worried about AI replacing their work: \textit{``Yes, the market needs to adjust heavily and I don’t think the revolution will be entirely peaceful''} (P2); \textit{``the idea of AI is mostly uncanny right now.''} (P4), and \textit{``Yes I [worry]. (...) a lot of tasks such as writing micro copy for websites etc which UX writers currently do would be automated''} (P14). Three more noted that they worry to some degree, or that they worry but are optimistic, e.g.: \textit{``I worry about it, but I hope the reality will be that AI becomes another tool''} (P5).

Nine respondents noted that they did not worry at all, while six reported that they do not worry \textit{yet}, e.g. \textit{``for now only the boring parts would be replaced. But this take-my-job-away argument was made countless times in history, there will always be something new. We can't be held back by this fear''} (P12). We group \textbf{reasons for concern} (aside from losing work) into the following themes. 

\vspace{-0.3em}
\noindent \paragraph{\textbf{1. Worse quality output.}} P8 observed: \textit{``It concerns me already that video games are becoming something of an echo chamber, and the sheer volume of games being released are diluting the market and making it harder for indie games to get the recognition they need to do well.''} The concern expressed here is not only that humans may become obsolete in the development process, but that the volume of output (in this case, of games) that AI (co-)creation makes possible will increase quantity but reduce quality of video games. 

P9 wrote \textit{``I certainly don't intend to replace all my hires with AI but some people will. They may achieve early success and they may also bring the genre into disrepute if they pump out a lot of lazy AI-written content.''} This indicates worries that extend beyond individuals and their job security to concerns about an entire genre of creative content. This perspective assumes that AI produces creative output of worse quality than humans produce, which we could consider a reason \textit{not} to worry about AI-generated content. However, in this case, the potential of such content to `dilute' or `bring into disrepute' a whole genre or field presents a threat or concern to some creatives.

\vspace{-0.3em}
\noindent \paragraph{\textbf{2. Weakening the creative process.}} Most respondents pointed out that humans will still be required in AI-facilitated creative processes or that the computer will simply help automate the `boring tasks.' However, a few also reflected on what that might mean to the creative processes, e.g., \textit{``I also don’t like the way AI image generators get you results instantly. They skip the creative process and just take you straight to the result… [...] that just overlooks a super important part of a creative process, which is exploration. And emergence, where stuff just kind of comes out of the process but you never imagined it would. Or happy accidents! In that sense I think AIs could actually lead to a stagnation in the history of creativity, if AI turns out to weaken the ``creative muscle'''} (P5). 

P11 further noted that \textit{``the meaning of `creative' seems to be increasingly twisted to mean merely `original/surprising,' and partly because there is a tendency for many to be unaware of the amount of creativity that my work involves. [...] A lot is being lost.''} This observation raises seminal questions similar to those raised in other fields where complex human thought processes have historically been replaced or at least disrupted, such as the introduction of calculators in algebra: How does it affect human cognition if computational processes take over (part of) our thinking? Will we lose our ability to use those parts of our brain, or will it simply free up cognitive reserve to consider new and more significant issues? 

\vspace{-0.3em}
\noindent \paragraph{\textbf{3. Copyright issues.}} Generative AI works only because a dataset exists that it can be trained on, and this raises new copyright issues, as P16 notes, \textit{``the ethical implications of AI stealing other people's work without credit [...] make me a bit wary.''} Many established artists have raised concerns about this issue since those whose art is currently visible on the internet lack means to opt out of image training databases  or otherwise control how their art is used \cite{cnnTheseArtists,waxyInvasiveDiffusion,Heikkila20222, sparkes2022ai}. Interestingly, this concern was  mentioned directly by only one respondent, suggesting that either it is not a matter that appears to be a threat to creatives we surveyed \textit{or} that they expect that a technological solution will emerge to address it; indeed, measures to protect intellectual properties of images, such as \textit{watermarks}, are currently being developed \cite{zhang2018protecting}.

\textbf{Reasons for \textit{not} worrying about generative AI} having a deleterious effect on their professions were described in three themes: 

\vspace{-0.3em}
\paragraph{\textbf{1. AI cannot produce output without human input.}} As described in Section \ref{creativitydefinitions}, several respondents questioned a computer's ability to produce truly original output. This was also described as a reason not to worry about AI replacing creative production or problem solving since human input is needed for datasets to be trained on and verified: \textit{``Being able to generate a Rothko at the click of a button is only possible because Rothko himself had original thoughts - that isn't creativity''} (P8), and \textit{``human input is still needed to verify and maintain AI's work''} (P22).

\vspace{-0.3em}
\paragraph{\textbf{2. AI output is not \emph{convincing}}.} Several respondents also noted that they do not find AI-generated output completely `convincing' or original: \textit{``I don't see any authentic or convincing AI in artistic fields at all''} (P8); \textit{``I know it can create pretty, but I don't think it can create ``Wow! I have never seen anything like it!''''} (P23); and \textit{``at the moment it's a tool that when used skillfully can create awesome images, but there still needs to be someone with creative taste and an eye for imagery at the helm. AIs also tend to generate `samey' images to me''} (P20). Although this theme resembles the preceding one (AI needs human input to produce output), which pertains more to requiring a human in the \textit{process} of creating and maintaining generative AIs, whereas the current theme critiques generative AIs' \textit{output}.

\vspace{-0.3em}
\paragraph{\textbf{3. My work/creative process is too complex for AI to imitate.}} Finally, several respondents observed that their work process is too complex for AI to replace it: \textit{``No, the complexity and dependencies is [sic] too high in my work''} (P21); \textit{``[I do not worry] for user interface design, there's so much to consider and think through that I can't see an AI making something fluid yet''} (P16). Particularly in processes of original problem solving and client communication, human cognition was described as indispensable: \textit{``Even the new code that they write is still going to be unoriginal in terms of problem solving''} (P15); \textit{``We work very closely with clients and our work requires a lot of thought process behind it. Our main product is communication ideas and solving problems visually. Often we can do that better with a scribble than a fancy looking piece of art. You can never ask the AI about the intention/thoughts/feelings behind the product''} (P10).

\vspace{-0.3em}
\subsection{Exciting Times Ahead!}
\vspace{-0.2em}
Thirteen respondents noted that they are more or less unequivocally excited about AI contributing to creative work in their profession (such as \textit{``Yes!''} or \textit{``Absolutely, exciting times ahead!''} (P12)). Four volunteered some version of ``yes \textit{and} no,'' e.g., \textit{``To some extent. I think some people will be able to use it in a nice way''} (P7). We grouped specific reasons for being excited about the advent and adoption of generative AI technology in creative professions into three themes:

\vspace{-0.3em}
\paragraph{\textbf{1. AI can raise productivity for the individual or for larger processes.}} Several respondents imagined AI being used to raise productivity, either in terms of individual efficiency (e.g., eliminating repetitive tasks and thus allowing creatives to focus on `more important' work): \textit{``there are things that are more efficient to leave to machines which should pair with things that humans will be better at for the foreseeable future.''} (P19)) or in terms of cultivating higher output rates by streamlining processes: \textit{``it would streamline many of the standard questions in the field''} (P1). 

\vspace{-0.3em}
\paragraph{\textbf{2. AI can offer inspiration.}} In fields that require creativity, it is perhaps not surprising that respondents highlighted using quickly generated output as a source of \textit{inspiration} in their creative process. Creative professionals often rely on readily available examples of design for inspiration \cite{herring2009getting}, and the availability of AI to generate innumerable novel examples was seen as a powerful opportunity for `opening up new solution spaces,' e.g.: \textit{``It will allow me to iterate through a much bigger possibility space''} (P12); and \textit{``it will make some work a lot easier/more efficient as you can try out different ideas in a very short amount of time''} (P6). In this role, AI is imagined to augment what we call the \textbf{divergent} parts of the creative process by offering examples and opening up novel and larger solution spaces \cite{dorst2001creativity}. 

\vspace{-0.3em}
\paragraph{\textbf{3. AI can lead to higher quality output.}}
Finally, some respondents highlighted the opportunity for AI to yield higher quality output, partially for the two reasons above (offering novel inspiration and freeing up time to work on tasks more central to the creative core), and partially due to qualities inherent in the AI itself: \textit{``Any creative work is better as a team effort and differences are a driving force. AI is very different and I want to work with them''} (P2); \textit{``it's a powerful tool that can enhance my work. [...] I can see it slotting into a step between browsing Pinterest for reference art and sketching my own stuff''} (P20). Two respondents also mentioned using AI for \textbf{convergent} parts of the creative process, for instance, decision making and evaluation: \textit{``It can augment decision making''} (P14); and \textit{``it opens up to possibilities to create new solutions and evaluate in new ways''} (P21), although specific ways for evaluation to occur were not described further. 

%\vspace{-0,7em}
\section{Discussion: Opportunities for participation}
\vspace{-0.2em}
Although complex, it seems prudent and timely to tackle the issue of how to encourage populations to participate in the development of AI more broadly \cite{birhane2022power}. We consolidate our preliminary analysis into four categories of potential focus for the design of participatory AI for 
creatives: (1) \textbf{Understanding AI}, \textbf{(2) Coping with AI}, \textbf{(3) Adapting to AI}, and (4) \textbf{Exploiting AI}. These categories align with the participatory design approach presented by Sanders \cite{sanders2002user} by considering what end-users \textit{know} (= understand AI), \textit{feel} (= coping with AI), \textit{do} (= adapt to AI), and \textit{dream} (= exploit AI). The categories offer a framework for engaging professional creatives in participatory AI design in a meaningful way. One could ask questions that align with the framework, e.g., ``How might we help future users \textit{understand} this technology'' or ``How might we help future users \textit{adapt} to new work flows?'' 

\vspace{-0.5em}
\subsection{Understanding AI}
\vspace{-0.2em}
Some survey responses identify a superficial understanding of the technical side of AI. This is acceptable, just as it is not a requirement of driving a car that one understands how the engine works. However, creatives will be better prepared to use AI as creativity support tools and design materials if they have a working understanding of the tools and their limitations \cite{dove2017ux, long2020ai,falk2022materializing}, particularly the level of \textit{agency} that computers can be ascribed (as we saw, no responses that demonstrated a deep level of technical understanding also portrayed the computer as having a high degree of agency). 

We suggest that facilitating a truthful \textit{understanding} of AI is the first step in empowering these users to co-create with AI technology. It is easy to brush this responsibility off as a creatives-only undertaking. However, we believe that AI developers share an ethical responsibility to make their systems accessible and explainable to a broader public, in line with the HCI research agenda for explainable, accountable and intelligible systems \cite{abdul2018trends}.

%Fra Trends and Trajectories for Explainable, Accountable and Intelligible Systems: An HCI Research Agenda:
%While researchers in the ML and AI communities are working on making their algorithms explainable, their focus is not on usable, practical and effective transparency that works for and benefits people. Given HCI’s core interest in technology that empowers people, this is a gap that we as a community can help to address, to ensure that these new and powerful technologies are designed with intelligibility from the ground up.
%Så jeg ser det som at HCI research agenda'en netop fokuserer på wider publics, ved at introducere usability etc og fokus på at empower people...
% Nanna: ja, men "Explainable AI" er ikke traditionelt en HCI-disciplin, men en machine learning-disciplin, så det betyder lidt at vi ikke fejlapproprierer den i vores felt.
%Jeanette: men det er vel lidt det HCI netop gør i mange tilfælde XD Tager noget udefra og får det til at passe ind i en HCI vinkel

\vspace{-0.5em}
\subsection{Coping with AI}
\vspace{-0.2em}
%Can we add something here which doesn't frame AI as inevitable and add something about refusing AI? As a way to cope with AI... (R3)
In the longer term, it is inevitable that AI-generated content of many kinds will be ubiquitous in most of our lives. How should we cope? We posit that creatives should 
%\sandy{I don't really understand what "hone their skills at estimating provenance" means; can you clarify?}
hone their skills in creating and in evaluating creativity. The responses to our survey suggest that they can recognize and celebrate indispensable human properties of creativity and art, e.g.: \textit{``human[-like] creativity is due to a combination of experiences and impressions that are connected in ways that are largely defined by human culture, and also feelings/sensations [...] that are mostly haphazard, and which AI don't have''} (P11). 
%Refer to lit. instead? R1
Sharing worries, excitement, and coping strategies --- including avoiding AI, see, e.g, \cite{kotakuArtistsProtest,zhang2018protecting, stablediffusionlitigationStableDiffusion} --- as well as celebrating what is uniquely creative about human approaches seems an important and achievable goal of designing participatory AI. 
We imagine a future where generative AI openly celebrates the sources from which its data are harvested, and where creators of generative AI include input from end-users in their design processes.

\vspace{-0.5em}
\subsection{Adapting to AI}
\vspace{-0.2em}
When photography was invented, artists adjusted their activities to focus less on realism and more on interpretation, whether through impressionism, abstraction, or surrealism (see, e.g., \cite{thecollectorPhotographyPioneered} for a more elaborate discussion of this). As writing, translation, paraphrasing and poetry become increasingly automated, professional writers and editors may become ``bosses to bots,'' instructing them on what to write, how to tailor material, and what to re-write when results do not meet professional or personal standards. 

Where by \textit{coping} we mean respectfully considering the new reality that these technologies bring about, by \textit{adapting} we suggest more comprehensive inclusion of creatives in the development of specific generative AI models. Several respondents shared excitement about the possibilities of using AI to help automate bureaucracy, repetitive tasks, and boring work. The responsibility of facilitating adaptation, however, does not fall only on creatives. By understanding creative needs and processes, generative AI developers may tailor AI systems to help specific professions and crafts in a way that is not only meaningful for creatives, but that may enhance the development of AI itself, similar to how PD was originally meant not only to improve information systems but also to  empower workers \cite{clement1993retrospective}.
%This reflect the origin of PD, which was not only ``the improvement of the information system, but also the empowerment of workers so they can codetermine the development of the information system and of their workplace'' \cite{clement1993retrospective}.

%Similar to hybrid AI-collaborations? And how PD strived to improve graphic workers' agency and ownership over their own field.

\vspace{-0.5em}
\subsection{Exploiting AI}
\vspace{-0.2em}
Photography changed what painters did, but it also opened up a field and a new profession: photographer. Technologies such as ChatGPT will change what writers do. Journalists are likely to spend more of their efforts on investigation and acquiring stories and less time on wordsmithing the reports on those stories. A mystery writer may give increased attention to plot features and less to the word-by-word narrative. Completely new tools and media may come out of the new AI technologies, including new types of creative jobs; as P3 notes, \textit{``the far future might include both 2D and 3D assets, generated in real time, as the player interacts with the experience [...] Experiences still need to be controlled, to ensure a good user experience. Therefore it would probably increase the number of creative/technical positions within game companies.''} (P3). 

We hypothesize that such technology can reach its full potential only if creative professionals truly participate in its development. AI has sometimes been described as ``a new shiny hammer in search of nails'' \cite{vardi2021will}, i.e., the technology or tool is being developed ahead of its specific purpose. We posit that if generative AI is developed with participation from creatives, there is a chance not only of better integration of AI in specific creative work practices, but also of leveraging creative competencies to imagine completely new avenues for these technologies.

\vspace{-0.5em}
\section{Conclusion and future work}
\label{conclusion}
\vspace{-0.2em}
The insights presented in this abstract illustrate some of the ways in which creative professionals speculate about and anticipate how AI may impact their creative work practices. Based on the insights, we encourage engaging creatives in the development of generative AI, both in developing concrete technology and in managing larger project issues as representatives of their peers, in line with the ideals of participatory design \cite{sanders2002user, bodker2022participatory}.
Pathways for developing more participatory AI should consider how creatives may better \textit{understand}, \textit{cope with}, \textit{adapt to} as well as \textit{exploit AI}.

\looseness-1 While the scope of our study is limited, we believe that both technology development and opinions towards AI are changing so quickly that it is relevant to share these  preliminary results. We hope they will spark discussions and inform future research into how to develop and use AI in a way that encourages and requires participation of the people who will be affected most by these technologies in the future.
Since most creative fields represented in our study are software-oriented, it is possible that the expressed views are more open and welcoming towards AI.
Future research should include a more evenly distributed representation from different creative fields as well as obtain richer data by conducting interview studies.
%Furthermore, while a survey study is a time-efficient and easily distributed method to elicit perspectives and attitudes from relevant respondents, future research should include interview studies to obtain richer data.

\looseness-1 Furthermore, the respondents came from different creative industries, and their everyday work lives may therefore not necessarily be impacted in the same ways by generative AI. 
%(for example, P11 worries that many of their work tasks may become automated, while P10 does not worry because they believe their work include complexities which cannot immediately be addressed by existing generative AI systems.) 
We have also not characterized how each individual's understanding of AI relates to, for instance, their level of worry or expectations since we believe this would require a larger participant group and deeper investigation.

Future work could categorize different creative industries and identify \textit{which} and \textit{how} specific work tasks within these industries may be impacted by generative AI, as well as investigate different ways to support these creative practices with AI.

\begin{acks}
This research has been supported by the VILLUM Foundation, grant 37176 (ATTiKA: Adaptive Tools for Technical Knowledge Acquisition) and by the Austrian Science Fund (FWF) [P34226-N]. 
\end{acks}

%%
%% The next two lines define the bibliography style to be used, and
%% the bibliography file.
\bibliographystyle{ACM-Reference-Format}
\bibliography{sample-base}

%%
%% If your work has an appendix, this is the place to put it.
\newpage
\appendix

\section{Respondents overview}

\begin{table}[H]
    \centering
    \renewcommand{\arraystretch}{1.1}%
    \scalebox{0.8}{
    \begin{tabular}{c|c|p{1.3cm}|c|p{12.8cm}}
        \textbf{P\#} & \textbf{Age} & \textbf{Country} & \textbf{Gender} & \textbf{``Please describe your profession and how creativity plays a role in your profession/work'' (slightly shortened for overview)}  \\ \hline 
        1 & 41-45 & USA & Female & \textbf{Librarian}, My work involves a lot of creative problem solving to address user needs. \\ \hline 
        2 & 36-40 & Denmark & Male & \textbf{Digital designer, game designer}, creative use of concepts and interfaces to tailor experiences and evoke emotions. \\ \hline
        3 & 31-35 & Denmark & Male & \textbf{Designing, programming and testing games}. Creativity is important in designing a game, since it helps the rest of the team to understand where the goal of the game is. \\ \hline
        4 & 36-40 & Turkey & Female & I am a \textbf{researcher} in academia and creativity plays a key role in terms of connecting dots on different topics, organize and elaborate my ideas clearly, be fluent on different subjects and flexible in terms of changing my perspectives when needed.\\ \hline
        5 & 41-45 & Denmark & Male & \textbf{3d artist / illustrator}. I work in a team tasked with coming up with new game concept prototypes. My day job involves a lot of  hands-on art creation, where different types of creativity is a big part - in concepting new stuff, or in problem solving, or simply in painting something\\ \hline
        6 & 21-25 & Germany & Female & \textbf{Media Designer}, I need to come up with creative ideas for campaigns and minigames on websites. \\ \hline
        7 & 31-35 & Denmark & Male & I'm a \textbf{game designer} / developer. I do everything from coding, drawing, animation and ui/ux design. Creativity plays a large role in everything I do \\ \hline
        8 & 26-30 & UK & Female & I work in \textbf{games production}, so am responsible for the scheduling and scope of PC and console videogame projects. There's a significant element of creative problem-solving in my job in terms of administration and organisation, as well as need for creativity in helping to create the games themselves\\ \hline
        9 & 41-45 & UK & Female & \textbf{video game writer / developer} \\ \hline
        10 & 31-35 & Denmark & Female & \textbf{Creative Producer}. I work with concept art and manage concept artists \\ \hline
        11 & 51-55 & Denmark & Female & I work as a \textbf{fixer, coordinator, and assistant}. Because my job is to make people shine, regardless of their intrinsic potential, I have to be extremely creative (and imaginitive and persuasive) in finding ways to accomplish this, both in terms of finding solutions and in convincing clients to adopt my suggestions.  \\ \hline
        12 & 36-40 & Germany & Male & \textbf{Software Engineer}, sometimes coding needs to be creative \\ \hline
        13 & 41-45 & Germany & Female & \textbf{Research Assistent} \\ \hline
        14 & 31-35 & UK & Male & I am a \textbf{computer scientist}. As any other researcher, much of my work involves using creative methods - designing experiments, writing papers, analysing data etc \\ \hline
        15 & 21-25 & USA & Female & \textbf{student} with an interest in art and design\\ \hline
        16 & 21-25 & USA & Non-binary & \textbf{Student in HCI and Design}, with the goal of becoming a user interface designer or similar. Creativity is necessary to solve design problems and create visuals to explain solutions. \\ \hline
        17 & 31-35 & Germany & Male & I \textbf{research computational creativity}, I am also an educator. \\ \hline
        18 & 41-45 & UK & Male & \textbf{Creative} \\ \hline
        19 & 31-35 & Denmark & Male & \textbf{Experience Design} for websites \\ \hline
        20 & 31-35 & Denmark & Male & By day I'm a \textbf{UI/UX artist}. In past I've worked as a concept artist. \\ \hline
        21 & 41-45 & Denmark & Male & \textbf{Project management and design}, finding new solutions on a daily basis... \\ \hline
        22 & 21-25 & Morocco & Male & I'm a \textbf{PhD student} working on extracting entities and relations from morphological descriptions. Creativity is crucial for my work especially in coming up with solutions, presenting the explainability behind the provided solution as well as for presenting it to non technical meeting. \\ \hline
        23 & 31-35 & Denmark & Female & \textbf{I teach simple programming and computer game development} to 11-13 year olds. They work through the design process, ideation, sketching, creating original pixel art and coding their own games. \\ \hline
    \end{tabular}}
    \caption{Overview of survey respondents}
    \Description{An overview of survey respondents, including participant number, age, residing country, gender and response to the question: Please describe your profession and how creativity plays a role in your profession/work. The responses have been slightly shortened due to space limitations.}
    \label{tab:participants}
\end{table}

\clearpage

\section{Graph of responses to ``What is AI?''}
\vspace{0.5cm}

\begin{figure} [h!]
    \centering \includegraphics[width=\textwidth]{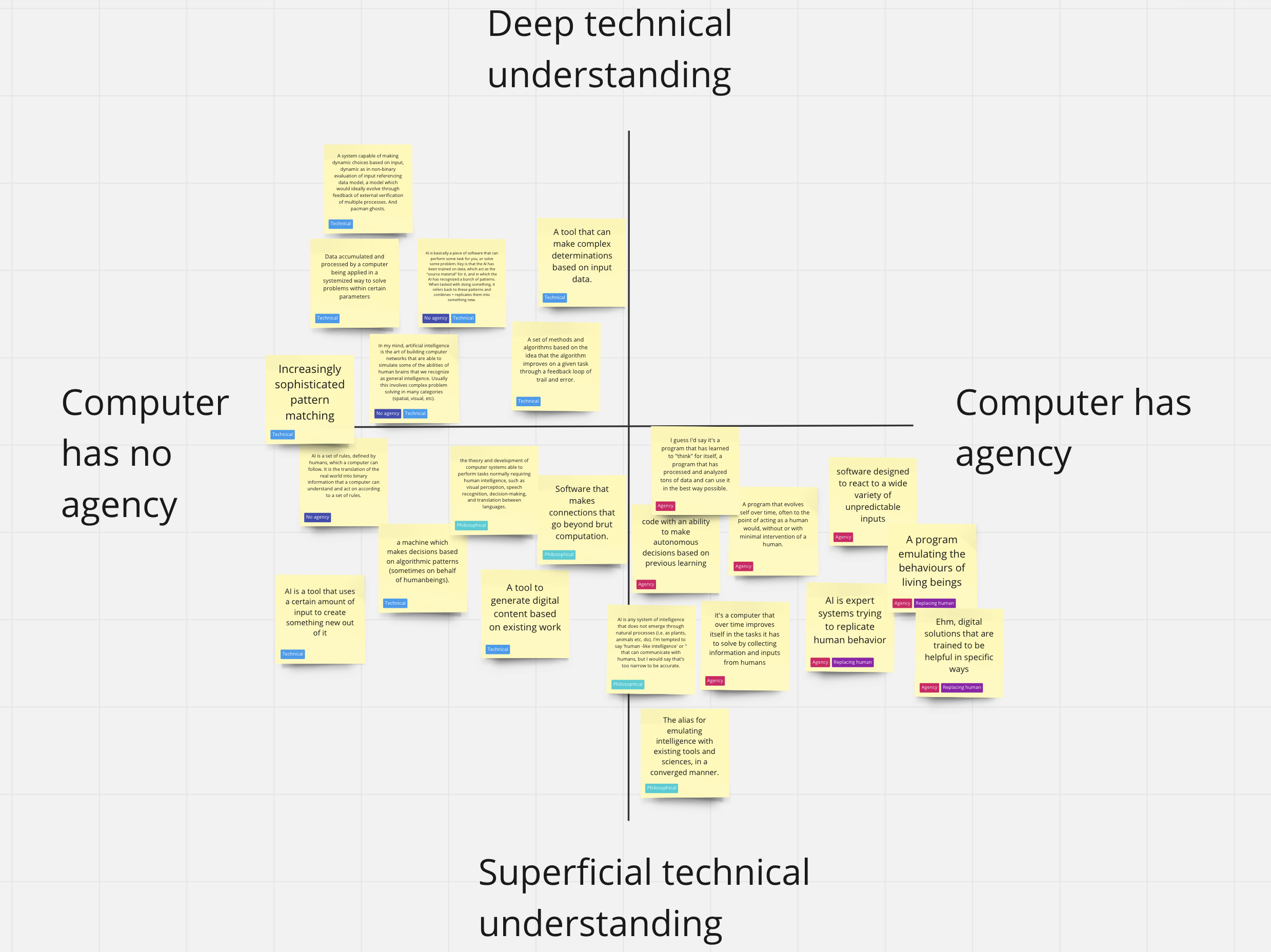}
    %\captionsetup{width=2\linewidth}
    \caption{Screenshot of the distribution of answers in terms of their technical depth and ascribed agency of AI.}
    \label{fig:whatisai}
\end{figure}

\end{document}